\documentclass
[twocolumn,aps,showpacs,floats,letterpaper,floatfix,groupedaddress,eqsecnum]{revtex4}%
\usepackage{amssymb,amsmath}
\usepackage[dvips]{graphicx}
\usepackage{dcolumn,epsfig}
\usepackage{amsmath}
\usepackage{amsfonts}
\usepackage{amssymb}%
\newcommand{\beq}{\begin{equation}}
\newcommand{\eeq}{\end{equation}}
\newcommand{\bes}{\begin{subequations}}
\newcommand{\ees}{\end{subequations}}
\newcommand{\bea}{\begin{eqnarray}}
\newcommand{\eea}{\end{eqnarray}}
\newcommand{\ba}{\begin{array}}
\newcommand{\ea}{\end{array}}
\newcommand{\beqn}{\begin{eqnarray*}}
\newcommand{\eeqn}{\end{eqnarray*}}

\newlength{\sizeonefig}
\newlength{\sizetwofig}
\begin{document}
\title{Dynamics of superconducting nanowires shunted with an external resistor}
\author{Matthew W.  Brenner$^{1}$, Dibyendu Roy$^{2}$, Nayana Shah$^{2}$, and Alexey Bezryadin$^{1}$}
\affiliation{$^1$University of Illinois at Urbana-Champaign, Department of Physics Urbana, Illinois 61801}
\affiliation{$^{2}$Department of Physics, University of Cincinnati, Cincinnati, Ohio 45221, USA}

\begin{abstract}
We present the first study of superconducting nanowires shunted with an
external resistor, geared towards understanding and controlling coherence and
dissipation in nanowires. The dynamics is probed by measuring the evolution of
the $V$-$I$ characteristics and the distributions of switching and retrapping
currents upon varying the shunt resistor and temperature. Theoretical analysis
of the experiments indicates that as the value of the shunt resistance is
decreased, the dynamics turns more coherent presumably due to stabilization of
phase-slip centers in the wire and furthermore the switching current
approaches the Bardeen's prediction for equilibrium depairing current. By a
detailed comparison between theory and experimental, we make headway into
identifying regimes in which the quasi-one-dimensional wire can effectively be
described by a zero-dimensional circuit model analogous to the RCSJ
(resistively and capacitively shunted Josephson junction)\ model of Stewart
and McCumber. Besides its fundamental significance, our study has implications
for a range of promising technological applications.

\end{abstract}
\date{February 20, 2012}
\maketitle


\section{Introduction}

\label{sec1}

The dissipation of the supercurrent in thin superconducting wires occurs solely  due to Little's phase slips \cite{Little67}. Advances in fabricating ultra-narrow superconducting nanowires has greatly boosted the interest in studying phase slippage in quasi-one-dimensional superconductors \cite{Bezryadin08}. There has been an intense activity to establish the
existence of quantum phase slips (QPS) related to macroscopic quantum
tunneling (MQT) \cite{Gio90, Bezryadin00, Golubev01, Lau01, Tian05, Alt06,
Zgirski08, Sahu09, LiWBBFC11} and to study quantum phase transitions between
possibly superconducting, metallic and insulating phases in
nanowires\cite{Bezryadin00, Lopatin05, Shah07, Refael07, DelMaestro08a,
DelMaestro08b, Bollinger08}. A dissipation-controlled quantum phase transition
\cite{Schmid83, Bulgadaev84, Chakravarty82, Zaikin97, Penttila99, Buchler04,
Werner05, Meidan07} has been predicted in junctions of superconducting
nanowires. Recently the importance of taking into account Joule-heating caused
by dissipative phase-slip fluctuations has also been argued and demonstrated
both theoretically and experimentally \cite{Tinkham03, Shah08, Sahu09,
Pekker09, LiWBBFC11}. Furthermore, quantum theory shows that the QPS rate as
well as the quantum phase transition can be controlled by an external shunt
\cite{Buchler04}. Dissipation plays an important role in dictating the physics
of nanowires. Conversely, superconducting nanowires provide an ideal prototype
for studying the interplay between coherence, dissipation and fluctuations.
Besides their fundamental importance, superconducting nanowires are also
ideally suited for building superconducting nano-circuitry and as devices with
potentially important applications, such as superconducting qubits and current
standards \cite{Mooij05, Mooij06}. Thus, even from the technological point of
view, it is extremely important to fundamentally understand the mechanism and
role of dissipation in nanowires and to find a way of experimentally
controlling coherence and dissipation. 

It is well established that the environmental dissipation of Josephson
junction (JJ) can be controlled by externally shunting the junction
\cite{Tinkham96}. This effect has been observed in the voltage-current
$(V$-$I)$ characteristics, which are greatly altered by the amount of
dissipation. The statistics of the switching and retrapping behavior in
shunted JJs have been investigated in last three decades and continues to be
actively studied \cite{BenJacob82, Martinis89, Vion96, Kivioja05, Krasnov07, Mannik05, Krasnov05, Myung09}. In general, the retrapping current, which is inversely proportional to the quality factor $Q$ of the circuit, is more sensitive to the amount of damping/dissipation than the switching current. The shunting is also known to control the rate of MQT of the phase variable in
superconductor-insulator-superconductor (SIS) junctions \cite{Takaqi05,
Leggett78, Leggett86, Caldeira81, Leggett87, Voss81, Martinis87, Martinis88}.
The Stewart-McCumber model \cite{Stewart68, McCumber68} of resistively and
capacitively shunted Josephson junctions (RCSJs) accurately describes much of
the physics of shunted JJs \cite{Tinkham96, Likharev81}. This model is quite
useful since it allows the analysis of various fundamental aspects of
superconducting devices, including chaotic behavior \cite{Khawaja08} and
high-frequency microwave responses \cite{Naito84}. The analysis of
superconducting computational circuits also involves use of the RCSJ model
\cite{Dimov05}.

Shunting a superconducting nanowire with a normal resistor should have a
strong effect on the superconducting character of the wire and just as in the
case of JJ could potentially provide a powerful way to control coherence and
dissipation. In spite of its clear importance and relevance, the behavior of
shunted nanowires has not been studied previously both experimentally and
theoretically. Here, we present the first study of shunted nanowires. It has
been inarguably proven for the case of unshunted nanowires that going beyond
linear response is essential to probe the dynamics of (quantum) phase-slip
fluctuations \cite{Tinkham03, Shah08, Sahu09, Pekker09,  LiWBBFC11}.
Furthermore, most applications would require the wire to be driven
out-of-equilibrium, making it doubly important to understand how the dynamics
of shunted nanowires\ evolves upon shunting. In fact, there is a third equally
important motivation for such a study. As discussed above, the RCSJ model has
been successfully used for JJs and has proven to be extremely important.
However, a circuit-element representation of a superconducting nanowire is
currently lacking and through this work we want to fill this gap by making
some concrete advances in that direction. 

The nanowires on which measurements were performed in the present work were
located in a low-pressure, thermalized helium gas and were fabricated using
the molecular templating method resulting in suspended nanowires
\cite{Bezryadin00, Bezryadin08}. It had already been demonstrated that these
superconducting nanowire show a large hysteresis in the $V$-$I$
characteristics for the unshunted case and that this hysteresis stems from
Joule heating and the strong temperature dependence of the resistance of the
wire \cite{Tinkham03, Shah08, Sahu09, Pekker09}. Local Joule heating by
phase-slip processes is especially important for a long free-standing nanowire
because the heat generated in the bulk of the wire is not removed easily and
has to flow away through the ends of the wire. Observation of similar physics
in a recent study of Aluminium nanowires fabricated using a different method
\cite{LiWBBFC11} points to the ubiquity and importance of Joule heating
effects and further underlines how it can be turned into an effective probe
for quantum phase slips. However, the best case scenario will be to be able to
have a control over the Joule heating. As will be shown in this paper heating
can indeed be controlled by shunting the superconducting nanowire with an
external resistance.

The article is organized as follows. In the next Sec. \ref{Samples} we briefly
describe the sample fabrication and measurement technique. The experimental
results are presented in Sec. \ref{Experiment}. This is followed by our
theoretical analysis and discussion in Sec. \ref{Theory} where we will argue
that shunting qualitatively changes the behavior of the nanowire and present
the theoretical results obtained by modelling the nanowire. Finally we will
end with concluding remarks in Sec. \ref{Conclusion}.

\section{Sample fabrication and measurement technique}

\label{Samples}

The nanowires presented in this study are fabricated using molecular
templating \cite{Bezryadin08, Bezryadin00}. Using electron-beam lithography
and a reactive ion etch, a 100 nm wide trench is patterned in the SiN layer of
a Si-SiO$_{2}$-SiN substrate. The trench is then etched in a $49\%$ solution
of hydrofluoric acid to form an undercut to prevent electrical leakage between
the electrodes, which are separated by the trench \cite{Bezryadin97}.
Fluorinated single-walled nanotubes, which are insulating, are dissolved in
isopropanol and then deposited onto the substrate containing the 100 nm wide
trench in the SiN layer and then dried with nitrogen gas. Randomly, some of
the nanotubes cross the trench, creating a scaffold for the nanowires to form
as the metal of choice is deposited on the substrate. The samples are then DC
sputtered with amorphous Mo$_{76}$Ge$_{24}$ in a high vacuum ($\sim10^{-7}$
Torr base pressure) chamber, thus coating the substrate and nanotubes with
12-18 nm of MoGe depending on the sample. A scanning electron microscope (SEM)
is then used to image the trench until a MoGe coated nanotube (nanowire) is
found to be relatively straight, homogeneous, and coplanar with the electrodes
\cite{Bezryadin08}. An SEM image of one such nanowire is shown in the inset of
Fig.\ref{pl2}(a). Contact pads are formed using photo lithography and wet
etching in a $3\%$ solution of H$_{2}$O$_{2}$, which etches MoGe rapidly.

All of the samples studied in this paper are $\sim$100 nm long and are
fabricated using MoGe. The thickness of each nanowire is controlled by the
deposition time in the sputtering chamber and by the configuration of
nanotubes used as a scaffold. The actual width of each sample is measured from
the SEM image and found to be $\sim$15, 12, 10, 15, 8, and 18 nm for samples
S1, S2, S3, S4, S5, and S6 respectively. Thicker samples show a lower normal
resistance $R_{\mathrm{n}}$, higher critical temperature $T_{\mathrm{c}}$,
higher critical current $I_{\mathrm{c}}$, and slightly higher retrapping
current $I_{\mathrm{r}}$. For example, for samples S1, S2 and S3 which have a
decreasing thickness, the resistance $R_{\mathrm{n}}$ in the normal state and
critical temperature $T_{\mathrm{c}}$ are S1 $(R_{\mathrm{n}}%
=1385~\mathrm{\Omega},~T_{\mathrm{c}}=4.607~\mathrm{K})$, S2 $(R_{\mathrm{n}%
}=1434~\mathrm{\Omega},~T_{\mathrm{c}}=4.41~\mathrm{K})$, and S3
$(R_{\mathrm{n}}=1696~\mathrm{\Omega},~T_{\mathrm{c}}=3.82~\mathrm{K})$. All
samples visually show a similar behavior in the $R$-$T$ and $V$-$I$ curves.

The shunt has been added by attaching a commercial metal film resistor
(ranging from 5 to 200 $\Omega$) parallel to the sample using silver paste
(Fig.\ref{pl1}(b)). The distance from the nanowire to the shunt is 1-2 cm for
all samples, and the shunt resistance is measured as a function of temperature
down to cryogenic temperatures and found to be constant. Measurements are
performed in a $^{4}He$ or $^{3}He$ cryostat equipped with base temperature
silver paste and copper powder filters and room temperature $\pi$-filters.
Transport measurements are carried out by current biasing the sample through a
large resistor ($\sim$1 M$\Omega$) and measuring the voltage with a
battery-operated Stanford SR 560 preamp, using the typical film-inclusive
four-probe technique as in Fig.\ref{pl1}(b) \cite{Bezryadin00}. Resistance vs.
temperature ($R$-$T$) curves are measured by applying a small sinusoidal
current ($\sim$10-100 nA) at a frequency of $\sim$12 Hz and measuring the
voltage and then doing a linear fit to the resulting voltage-current data to
obtain the resistance. The temperature is measured using a calibrated Cernox
thermometer from LakeShore. $V$-$I$ curves are measured by applying a large
sinusoidal current in the range of a few $\mathrm{\mu}$A, at a frequency of a
few Hz, and measuring the voltage simultaneously. The switching (retrapping)
current has been measured by sweeping the current as in the $V$-$I$
measurement, and recording the current at which the voltage-jump (drop) out of
the superconducting (resistive) state has been the greatest.

\section{Experimental results}

\label{Experiment}

In this section we present a series of experimental results. We begin by
presenting the temperature dependence of the linear resistance of shunted
nanowires and use that as a benchmark for characterizing different samples.
Next, we go one step further and present the measurements of the nonlinear
$V$-$I$ characteristics of the current-biased nanowires at temperature of
$1.8$ K, focusing on the evolution of hysteresis upon shunting. In the next
subsection the experimental data of switching and retrapping distributions for
different values of shunt are presented. Such an analysis provides deeper
insights into the dynamics of superconducting nanowires as is evident from a
previous study in the case of unshunted wires\cite{Shah08, Sahu09, Pekker09}.
Although the focus of our work is on studying the effect of shunting, in the
last subsection we present the temperature dependence of the $V$-$I$
characteristics and the switching distributions so as to provide a comparison
with the corresponding measurements for the unshunted case that was studied in
detail\cite{Shah08, Sahu09, Pekker09}.

\subsection{Shunt dependence of $R(T)$: characterization of the samples}

\begin{figure}[ptb]
\includegraphics[width=8cm]{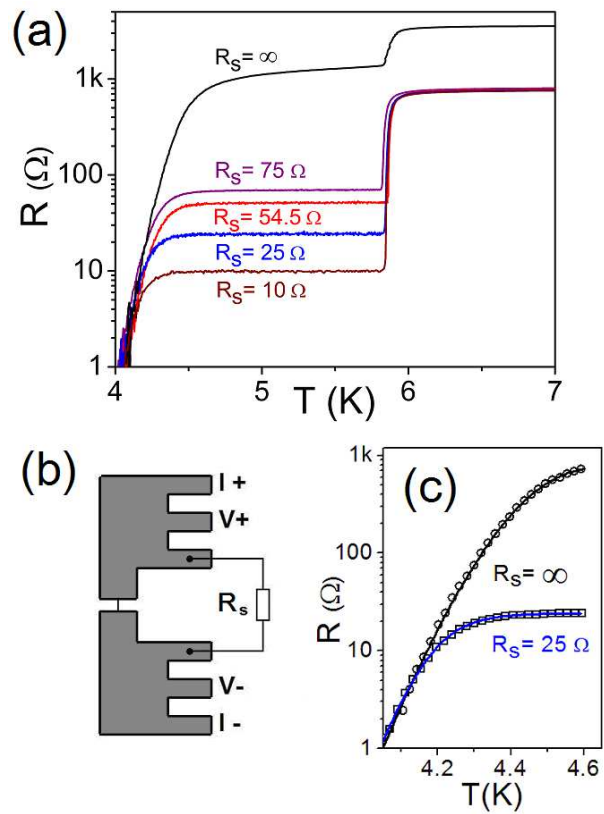}\caption{(a) $R$-$T$ data for sample S1
with various shunt values. The first sharp drop in resistance at ~5.8 K is due
to the film electrodes going superconducting. The second, gradual, drop in
resistance at lower temperature is the superconducting transition of the
nanowire, which is much broader due to TAPS. (b) Sample schematic. The wire is
shown as a short vertical line and is shunted by a commercial resistor
$(R_{\mathrm{S}})$. The sample is measured by current biasing the sample and
extracting the resistance via a four-probe measurement. (c) Comparison of
$R$-$T$ curves from (a) for the case of no shunt (circles) and a $25~\Omega$
shunt (squares) with the theoretical expression of the total sample resistance
given by Eq.(\ref{tres}) (solid lines). The known parameters used in each fit
are: wire length, L=105 nm, normal state resistance, $R_{\mathrm{n}%
}=1385~\Omega$, and shunt resistance, $R_{\mathrm{S}}=\infty$ and
$R_{\mathrm{S}}=25~\Omega$, respectively. The fitting parameters used for the
fits are $T_{\mathrm{c}}=4.607~K$ for the unshunted case and $T_{\mathrm{c}%
}=4.595~K$ for the $25~\Omega$ shunted case, and $\xi(0)$=7.57 nm for both
cases, where $\xi(0)$ is the dirty limit coherence length at zero
temperature.}%
\label{pl1}%
\end{figure}Study of the linear-response resistance as a function of
temperature is useful for characterizing the samples and establishing a
starting point for further investigation. Fig.\ref{pl1}(a) shows the
temperature dependence of the nanowire's resistance using a log-linear scale
for various values of the shunt resistance $R_{\mathrm{S}}$. As the
temperature is lowered below 5.8 K the film becomes superconducting while the
wire is still resistive because its critical temperature $(T_{\mathrm{c}})$ is
lower than that of the film. Below $T_{\mathrm{c}}$ of the wire, as expected
there is a measurable resistance due to phase slips in the wire.

We understand the measured resistance vs. temperature $(R$-$T)$ curves using
the following arguments. Below $T_{\mathrm{c}}$ of the nanowire the total
sample resistance is a parallel combination of the $R_{\mathrm{S}}$ and the
wire resistance, $R_{\mathrm{W}}$. We model the wire resistance with an
empirical formula,
\begin{equation}
\frac{1}{R_{\mathrm{W}}(T)}=\frac{1}{R_{\mathrm{n}}}+\frac{1}{R_{\mathrm{AL}%
}(T)}\label{wireRes}
\end{equation}
where $R_{\mathrm{n}}$ is the normal state resistance of the nanowire to
account for the quasi-particle resistance channel and $R_{\mathrm{AL}}$ is the
Arrhenius-Little (AL) resistance occurring due to thermally activated phase
slips (TAPS). The AL resistance is estimated, following Little's proposal, by
assuming that each phase slip creates a normal segment on the wire of a size
equal to the coherence length and for a time interval roughly equal to the
inverse attempt frequency \cite{Little67, Bezryadin08}. We note that the
Langer-Ambegaokar-McCumber-Halperin (LAMH) theory \cite{Langer67, McCumber70} of TAPS is not
valid except very near to $T_{\mathrm{c}}$ \cite{Meidan07}. So we have to use
the phenomenological AL expression:
\begin{align*}
R_{\mathrm{AL}}(T) &  =R_{\mathrm{n}}~\mathrm{exp}\Big(-\frac{\Delta
F(T)}{k_{\mathrm{B}}T}\Big),\\
\mathrm{where}~~\Delta F(T) &  =\frac{8\sqrt{2}}{3}\Big(\frac{H_{\mathrm{c}%
}^{2}(T)}{8\pi}\Big)A\xi(T)
\end{align*}
is the free energy barrier for a phase slip in the zero-bias regime
\cite{Langer67}. Here $H_{\mathrm{c}}(T)$ is the thermodynamic critical field,
$\xi(T)$ is the temperature dependant coherence length, $A$ is the
cross-sectional area of the wire, and $k_{\mathrm{B}}$ is the Boltzmann
constant. The equation for the free energy barrier $\Delta F(T)$ can be
rewritten to include wire parameters more accessible via the experiment as
\cite{Tinkham02}
\begin{equation}
\Delta F(T)=0.83k_{\mathrm{B}}T_{\mathrm{C}}\Big(\frac{R_{\mathrm{Q}}%
}{R_{\mathrm{n}}}\Big)\Big(\frac{L}{\xi(0)}\Big)\Big(1-\frac{T}{T_{\mathrm{C}%
}}\Big)^{3/2},
\end{equation}
where $L$ is the length of the nanowire and $R_{\mathrm{Q}}(=h/4e^{2}%
\approx6450~\Omega)$ is the quantum resistance. Thus, the
temperature-dependent total sample resistance is:
\begin{equation}
R_{\mathrm{T}}(T)=\Big[\frac{1}{R_{\mathrm{S}}}+\frac{1}{R_{\mathrm{n}}%
}\Big(1+\mathrm{exp}\Big(\frac{\Delta F(T)}{k_{\mathrm{B}}T}%
\Big)\Big)\Big]^{-1} \label{tres}
\end{equation}
The fits of the total sample resistance by Eq.(\ref{tres}) are presented in
Fig.\ref{pl1}(c) for the unshunted nanowire $(R_{\mathrm{S}}=\infty)$, and the
case when the nanowire is shunted with $25~\Omega$. All the fits are done by
using the values of $R_{\mathrm{n}}$, $T$, $L$, and $R_{\mathrm{S}}$ obtained
from the $R$-$T$ curve and the SEM image and using $T_{\mathrm{c}}$, and
$\xi(0)$ as fitting parameters. The value of the fitting parameters change
very slightly as the shunt resistance is varied. For instance, in the fitting
presented in Fig.\ref{pl1}(c), the value of $T_{\mathrm{c}}$ decreased by 12
mK for the shunted case compared to the unshunted case, which can be accounted
for by slight sample change during thermal cycling. The agreement between
experimentally measured resistance and the total resistance in Eq.(\ref{tres}%
), as shown in Fig.\ref{pl1}(c), gives evidence that the $T$-dependence of the
rate of phase slips does not depend on the shunt at relatively high
temperatures ($T>4~K$ in this case) and that the observed residual resistance
of the wire just below $T_{\mathrm{c}}$ is due to TAPS in the high temperature
limit of $\sim4-5~K$. Note also that the $R$-$T$ curves of all the samples
presented in this paper are smooth and show no extra transitions, and the SEM
images confirm that the nanowires are homogeneous and well-connected to the
electrodes. Thus our nanowires are well suited to systematically study the
effect of shunting.

\subsection{Shunt dependence of $V$-$I$\ characteristics}

\begin{figure}[ptb]
\includegraphics[width=8cm]{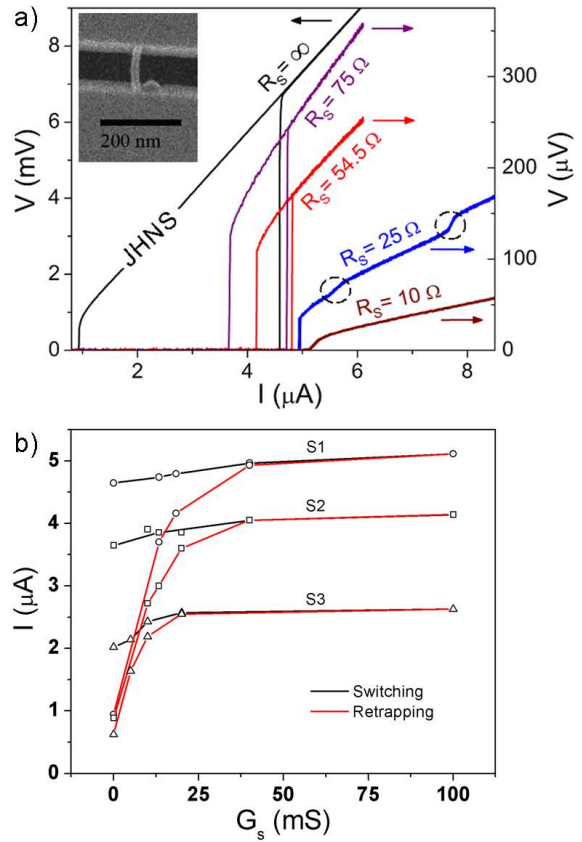}
\caption{(a) Voltage vs. total current for sample S1 at 1.8 K corresponding to
various values of the shunt resistance, $R_{\mathrm{S}}$. Dashed circles are
kinks in the V-I curve occuring for the case when the nanowire is shunted with
$R_{\mathrm{S}}=25~\mathrm{\Omega}$. Inset: SEM image of the nanowire for
sample S1. (b) Mean switching and retrapping currents vs. $G_{\mathrm{s}%
}~(\equiv1/R_{\mathrm{S}})$ for three samples (S1 at 1.8 K, S2 at 1.5 K, and
S3 at 1.6 K).}%
\label{pl2}%
\end{figure}

Let us start by considering the unshunted wire $(R_{\mathrm{S}}=\infty)$. As
the bias current $I$ is increased, thermal fluctuations cause the nanowire to
switch from a superconducting state into a resistive state before the current
reaches the critical (equilibrium) depairing current. The current at which the
wire switches out of the superconducting state is called the switching current
$(I_{\mathrm{sw}})$. Once in the resistive state, as the current is decreased
below some critical value of current, the nanowire experiences retrapping back
into the superconducting state. The current at which this happens is called
the retrapping current $(I_{\mathrm{r}})$. For unshunted wires, switching is
stochastic in nature, i.e., each new current sweep gives a different value for
the switching current but the retrapping process is non-stochastic. We devote
the next subsection for the discussion of switching and retrapping
distributions and their dependence on the shunt resistance. Here we focus on
the mean values of the switching and retrapping currents and show how they
evolve upon shunting the wire as shown in Fig.\ref{pl2}. 

Fig.\ref{pl2}(a) shows the $V$-$I$\ characteristics for different values of
shunt resistance. As the nanowire is shunted with lower values of the shunt
resistance, the mean switching and retrapping currents are increased while the
width of the hysteresis is decreased. Additionally, the retrapping current
also becomes stochastic (as shown in Fig.\ref{rpEdist}). In Fig.\ref{pl2}(b),
the dependence of the mean switching and retrapping currents on the shunt
resistance are shown for different nanowire samples. The mean switching
current increases, at a lower rate than the retrapping current, and saturates
for small values of the shunt resistance (Fig.\ref{pl2}(b)) with a decreasing
(increasing) shunt resistance $R_{\mathrm{S}}$ (conductance $G_{\mathrm{S}})$.
Similarly, the retrapping current increases with decreasing the shunt
resistance $R_{\mathrm{S}}$ until $I_{\mathrm{r}}$ finally reaches
$I_{\mathrm{sw}}$ of the wire. Such behavior is observed on all tested
samples. As the switching and retrapping current coincide, the hysteresis
disappears. Nanowires with smaller $T_{\mathrm{c}}$ start showing this
saturation behavior at higher shunt values.

When the nanowire is shunted with a $25~\mathrm{\Omega}$ resistor or less,
kinks in the voltage are also observed (they are marked by dashed line circles
in Fig.\ref{pl2}(a) for the case where sample S1 is shunted by
$25~\mathrm{\Omega}$. The kinks for the $10~\mathrm{\Omega}$ shunting case are
not shown, but occur at higher current. We relegate the interpretation of
these kinks to Subsec. \ref{VIsim}.

\subsection{Shunt dependence of switching and retrapping distributions}

\begin{figure}[ptb]
\begin{center}
\includegraphics[width=8cm]{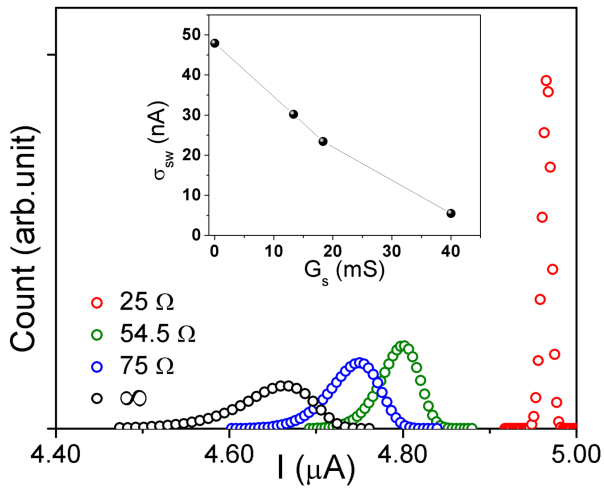}
\end{center}
\caption{Switching distributions vs. bias current for sample S1 shunted with
various $R_{\mathrm{S}}$ values at 1.8 K. All distributions were measured with
a sinusoidal current sweep with $f=8~\mathrm{Hz}$ and an amplitude of 6.1
$\mathrm{\mu A}$. Inset: Shows the standard deviation $\sigma_{\mathrm{sw}}$
of switching current distribution as a function of $G_{\mathrm{s}}%
~(\equiv1/R_{\mathrm{S}})$.}%
\label{pEdist}%
\end{figure}As mentioned in the previous subsection, the unshunted nanowire
undergoes stochastic switching as the bias current is increased. We plot the
switching distributions vs. current in Fig.\ref{pEdist} for different values
of external shunt resistance to see how it evolves upon shunting. Lowering the
value of the shunt resistance has the effect of narrowing the width,
increasing the height, and shifting the distribution to higher currents. As
can be seen from the plots, the full width of the distribution at half maximum
(FWHM) changed from 100 nA to 12 nA due to shunting with a $25~\mathrm{\Omega
}$ resistor. The asymmetric shape in the distribution for larger shunts
changes to a rather symmetric shape with lower shunts.

\begin{figure}[ptb]
\begin{center}
\includegraphics[width=8cm]{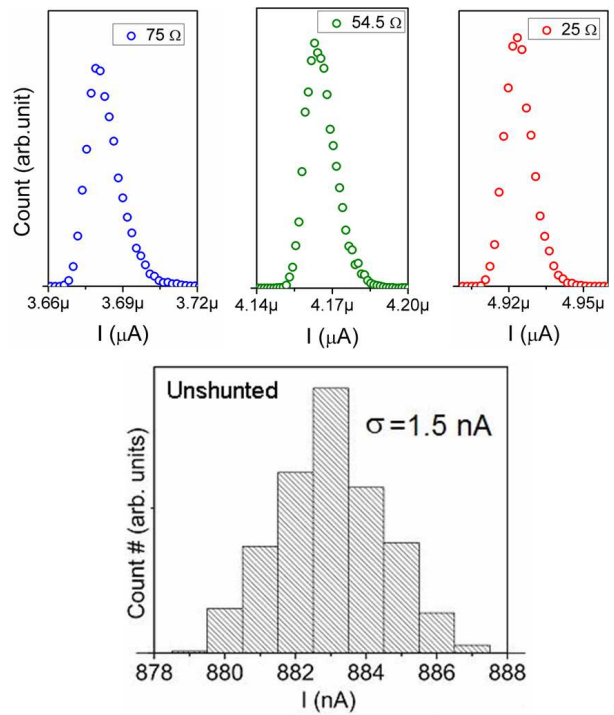}
\end{center}
\caption{Top: Retrapping distributions vs. bias current for sample S1 shunted
with various $R_{\mathrm{S}}$ values. The mean and standard deviation of the
retrapping data are: $3.683,4.166,4.925~\mathrm{\mu A}$ and $7.304,6.089,$ and
$5.799~\mathrm{nA}$, respectively for $R_{\mathrm{S}}%
=75,54.5,25~\mathrm{\Omega}$. Bottom: Typical retrapping histogram vs. $I$ for
the case where the nanowire is unshunted. The standard deviation of the
retrapping current of the unshunted wire matches with the experimental current
noise.}%
\label{rpEdist}%
\end{figure}

The retrapping current also shows a dramatic change from deterministic values
to stochastic values when shunted with 75 $\mathrm{\Omega}$ or less. The
bottom part of Fig.\ref{rpEdist} is a typical retrapping histogram for an
unshunted wire\cite{Sahu09}. Here, the standard deviation of the retrapping
current is 1.51 nA, which is the noise limit of our experimental setup. So,
this small distribution of retrapping current is just due to the instrumental
noise, and can be reduced by decreasing the instrumental noise and the spacing
in between bias-current points. Thus, retrapping in unshunted nanowire always
occurs at the same current, i.e. the transition is deterministic. However,
when the wire is externally shunted, a retrapping distribution is observed,
with its width being much larger than the experimental setup noise and
independent of the bias-current spacing. 

In the top of Fig.\ref{rpEdist}, the retrapping current distributions for
sample S1 are shown for the nanowire shunted with different values of external
resistances. The width of the distribution is slightly sensitive to the value
of shunt resistance, but the mean value of the retrapping current changes
considerably. When shunted with $75~\mathrm{\Omega}$ for instance, the
standard deviation of the retrapping current increases above the experimental
setup noise to 7.3 nA (from ~1.5 nA for the unshunted case under the same
conditions). Interestingly the width of the retrapping and switching current
distributions for 25 $\mathrm{\Omega}$ shunt is almost similar. The retrapping
distributions for the shunted nanowire are asymmetric in contrast to the
unshunted case in which the distribution is symmetric as can be seen in the
bottom of Fig.\ref{rpEdist}.

\subsection{Temperature dependence: shunted vs. unshunted nanowires}

\begin{figure}[ptb]
\begin{center}
\includegraphics[width=8cm]{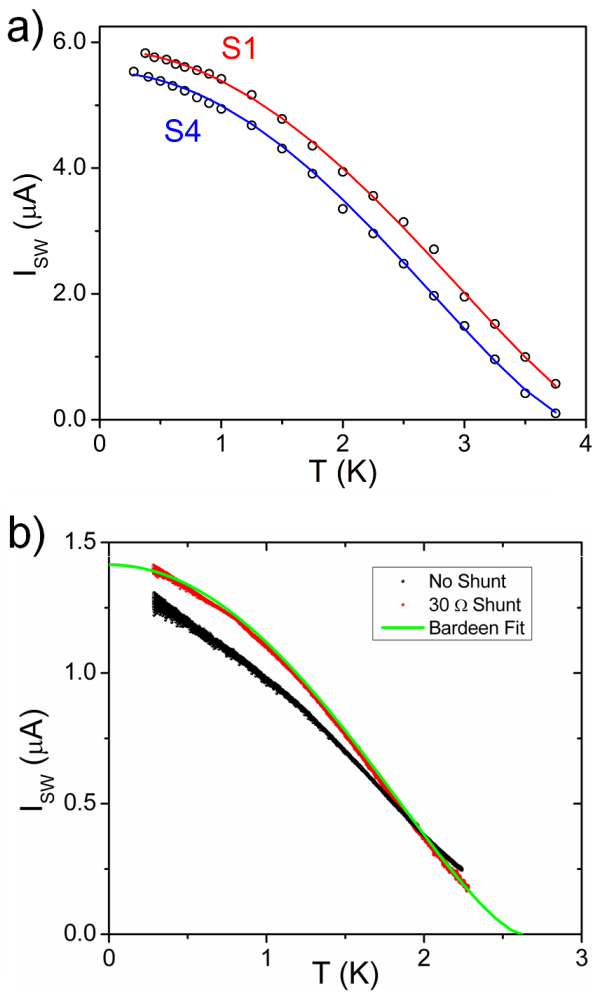}
\end{center}
\caption{(a) Mean $I_{\mathrm{sw}}$ vs. $T$ for samples S1 and S4 shunted with
$5~\mathrm{\Omega}$ and $10~\mathrm{\Omega}$ respectively. The solid lines are
fits to the Bardeen's prediction (Eq.\ref{Bardeen}) for the temperature
dependence of the depairing current, $I_{\mathrm{c}}$. The fitting parameters
used are: $I_{\mathrm{c0}}=5.88,~5.53~\mathrm{\mu A}$ and $T_{\mathrm{c}%
}=4.2,3.9~$K for samples S1 and S4, respectively. (b) Distribution of
$I_{\mathrm{sw}}$ vs. $T$ for sample S5 unshunted (red) and shunted with 30
$\mathrm{\Omega}$ (black). Each curve contains approximately 20,000 points.
The corresponding fit to the Bardeen prediction (green) is presented for the
shunted wire. Here $I_{\mathrm{c0}}=1.415~\mathrm{\mu A}$ and $T_{\mathrm{c}%
}=2.62~K$.}%
\label{dep}%
\end{figure}In this subsection we discuss the temperature evolution of the
dynamics of shunted nanowires. In Fig.\ref{dep}(a), the mean value of the
switching current is plotted at various temperatures for sample S1 shunted
with $5~\mathrm{\Omega}$ and sample S4 shunted with $10~\mathrm{\Omega}$. In
Fig.\ref{dep}(b), a distribution of switching currents is plotted as a
function of temperature for sample S5 when it is shunted with a
$30~\mathrm{\Omega}$ resistor and compared with the unshunted case. As the
temperature is reduced, the switching current for all samples increases and
begins to show signs of saturation below $1$~K. The behavior of the switching
current as a function of temperature in the distribution measurement in
Fig.\ref{dep}(b) is similar to that of Fig.\ref{dep}(a) except that in
Fig.\ref{dep}(b) the fluctuation in the switching current is also displayed.

To check the proximity of the switching current to the equilibrium depairing
current, Bardeen's prediction \cite{Bardeen62} for the temperature dependence
of the equilibrium critical (depairing) current is compared to the temperature
dependence of the measured switching current for nanowires shunted with small
resistances as shown in Fig.\ref{dep}. The Bardeen's equation, derived from
BCS theory, is given by:
\begin{equation}
I_{\mathrm{c}}(T)=I_{\mathrm{c0}}\Big(1-\big(\frac{T}{T_{\mathrm{c}}}%
\big)^{2}\Big)^{3/2} \label{Bardeen}
\end{equation}
where $I_{\mathrm{c0}}$ is the critical current at zero temperature. Excellent
agreement is found with the experimental data over a wide temperature interval
suggesting that the shunt has driven the switching current close to the
depairing current. In these fits, the temperature is known while the critical
current at zero temperature $I_{\mathrm{c0}}$ and the critical temperature
$T_{\mathrm{c}}$ are used as fitting parameters. Close agreement is found
between the theoretical prediction for the depairing current at zero
temperature: $I_{\mathrm{c0}}=92\mathrm{\mu A}(LT_{\mathrm{c}})/R_{\mathrm{n}%
}\xi(0)$ \cite{Tinkham02}, which is derived from BCS and Ginzburg-Landau
theory, and the value of $I_{\mathrm{c0}}$ used in the Bardeen fit. Here $L$
and $\xi(0)$ are in nm, $T_{\mathrm{c}}$ in K , and $R_{\mathrm{n}}$ in
$\mathrm{\Omega}$. Using the fitting parameters from the $R_{\mathrm{T}}(T)$
fit of Eq.(\ref{tres}) presented in Fig.\ref{pl1}(c), $I_{\mathrm{c0}}$ has a
theoretical value of $5.48~\mathrm{\mu A}$, while $I_{\mathrm{c0}}$ used in
fitting to the Bardeen formula in Fig.\ref{dep} has a value of
$5.88~\mathrm{\mu A}$. Thus, excellent agreement is found between the
theoretical and experimental value for $I_{\mathrm{c0}}$. A value of $4.2~$K
for $T_{\mathrm{c}}$ is used to fit the temperature dependence of the
switching current with the Bardeen formula, while the $R_{\mathrm{T}}(T)$ fit
using the AL model predicts the value of $T_{\mathrm{c}}$ to be $4.717~$K.
This difference can be accounted for by sample oxidation and thermal cycling
between measurements.

\begin{figure}[tbh]
\includegraphics[width=8cm]{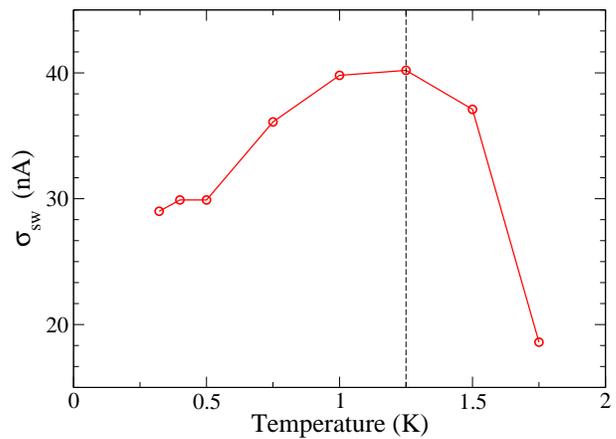}
\caption{The standard deviation $\sigma_{\mathrm{sw}}$ vs. $T$ for the
switching events of sample S6 shunted with $10~\mathrm{\Omega}$, where for
each point $\mathrm{\sigma}$ was calculated using data sets of 10000 points.
}%
\label{swwd1}%
\end{figure}Finally in Fig.\ref{swwd1}, we plot the measured temperature
dependence of the switching distribution width for the shunted nanowires. We
find that it shows a trend similar to that previously observed for the
unshunted wire\cite{Sahu09}. As for the retrapping current, which is observed
to be stochastic only in the shunted nanowire, the standard deviation is never
observed to increase with decreasing temperature. However, at low
temperatures, MQT is expected to cause the standard deviation of the
retrapping current to be constant with temperature \cite{Chen88}. In our
experiments, we have seen evidence of this behavior and will investigate this
more fully in the future.

\section{Theoretical analysis and discussion}

\label{Theory}In this section we will develop a theoretical interpretation and
understanding of the experimental data presented in the previous section. In
doing so, we will start by giving a brief account of the unshunted nanowires
which have been previously investigated in detail. Then we will argue how
shunting the nanowires brings about qualitative changes in the dynamics and
develop a physical picture by gathering experimental signatures and
theoretical arguments. Next we will motivate the theoretical model for
explicit calculations and numerical simulations and use it to generate the
$V$-$I$ characteristics and the distributions that will be compared with
experimental measurements.

\subsection{Unshunted nanowires}

Recently, properties such as the $V$-$I$ characteristics and the switching
distributions of the unshunted nanowire have been studied in detail to
understand the behavior of quasi-one-dimensional superconductors at low
temperatures \cite{Shah08, Sahu09, Pekker09, LiWBBFC11}. In
quasi-one-dimensional superconductors the zero resistance superconducting
state is destabilized by thermal and quantum phase-slip fluctuations. These
phase-slip fluctuations induce resistance which causes Joule heating in the
nanowire. If this heat generated by phase-slip fluctuations in the bulk of the
wire is not overcome sufficiently rapidly, it can reduce the depairing current
to below the applied current, thus causing transition to the highly resistive
state. It has been found in experiments with unshunted nanowire that while the
distribution of re-trapping currents is very narrow and almost
temperature-independent, the distribution of switching currents is relatively
broad and the mean as well as the width of the distribution change with
temperature of the leads. The distribution in switching currents reflects that
the collective dynamics of the superconducting condensate evolves
stochastically in time and undergoes phase slip events at random instants. A
stochastic model for the time-evolution of the temperature in a nanowire has
been developed to understand above experimental results \cite{Shah08,
Pekker09}. The model predicts that although, in general, switching from
superconducting to resistive normal state occurs due to several phase-slip
events, it can be even induced by a single phase slip at a particular
temperature and current-range. The model also indicates non-monotonic
temperature dependence of the width of the distribution of switching currents.
Thus, these experiments with switching events as well as those with microwave
radiation in unshunted wires suggest that the resistive state is the normal
state of the wire maintained by Joule heating, i.e. the JHNS \cite{Tinkham03,
Sahu09, Dinsmore08, LiWBBFC11}. For unshunted wires, the retrapping process is
non-stochastic since the retrapping occurs from the thermalized Joule heating state.

\subsection{Qualitative picture of shunt-induced crossover}

\label{QualitativePicture}

\begin{figure}[tbh]
\includegraphics[width=8cm]{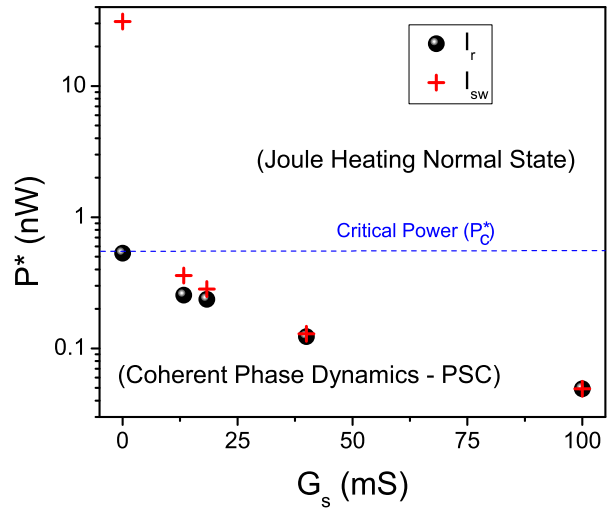}
\caption{Log-linear plot of average power dissipated in the nanowire
immediately after a switching event (crosses) and just before retrapping
(circles), plotted versus the shunting conductance $(=1/R_{\mathrm{S}})$.
Above a critical power, $P^{*}_{\mathrm{c}}$, the resistive state of the
nanowire is identified as the JHNS while below $P^{*}_{\mathrm{c}}$, the
resistive state originates from a coherent phase dynamics. The power is
calculated by taking the voltage across the wire times the current through the
wire $(P^{*}=I_{\mathrm{W}}V)$ at the retrapping and switching current in the
resistive branch of the $V$-$I$ curve from Fig.\ref{pl2}(a). }%
\label{Cpow}%
\end{figure}

How does the picture for the unshunted case discussed above evolve as we shunt
the wire with an external resistance? With the inclusion of a shunt resistance
the applied bias current is divided into two parts, and the part going through
the nanowire decreases as the shunt resistance becomes smaller. A lower
(higher) current through the wire causes a decrease (increase) in Joule
heating in the wire and hence a decrease (increase) in local temperature. For
the unshunted case that corresponds to an infinite shunt resistance, the
heating is maximum and as discussed in the previous subsection one gets a JHNS
in the unshunted wire. 

We will argue that upon decreasing the value of the shunt resistance the
nature of the resistive state of the nanowire changes from a JHNS to a phase
slip center (PSC) state. PSC is a process of periodic-in-time phase rotation
occurring in a certain region of the wire and is driven by the bias current.
An ideal PSC acts qualitatively like a weak-link JJ in series with the rest of
the wire, and the differential resistance associated with it is determined by
the quasiparticle diffusion length \cite{Skocpol74}. This resistance is much
smaller than the normal resistance of the wire. This is what we indeed obtain
when the nanowires are shunted with small resistance. The most explicit proof
for a PSC would of course be the observation of Shapiro steps, because they
prove that there is a periodic phase rotation in the system. However we have
compelling arguments and consistency in our explanation that points to the
existence of PSC. 

The deterministic retrapping current observed for the unshunted nanowires
becomes stochastic when an external resistive shunt is added (at least within
the experimental range of shunt resistances). The deterministic retrapping
current reflects that the resistive state of the unshunted wires is a
thermalized JHNS. There is overheating and the wire is normal. As $I$ is
reduced the temperature goes down. Relative fluctuations of the temperature
are small since it is determined by a macroscopic number of normal electrons.
In this case $I_{\mathrm{r}}$ is fixed by the current at which the heating is
not enough to keep the system above $T_{\mathrm{c}}$. If the system must
re-trap at $I_{\mathrm{r}}^{0}$ then as can be seen in Fig.\ref{rpEdist}, the
distribution is symmetric and Gaussian, centered around $I_{\mathrm{r}}^{0}$,
and as wide as the noise in the bias current circuit and in the measurement
circuit can smear it. On the other hand, the stochastic retrapping current
indicates that the finite-voltage/resistive state of the shunted wires is
governed by a coherent dynamics of the phase of superconducting order
parameter. The dynamic state, called PSC moves by inertia, which is the
voltage on the electrodes. But a strong thermal (or quantum) fluctuation can
re-trap the system from the dynamic to the static state. And such change will
be permanent. So the retrapping can happen at $I>I_{\mathrm{r}}^{0}$. The
distribution is asymmetric (right-skewed or right-tailed) since the system can
never switch at $I<I_{\mathrm{r}}^{0}$. Here, the fluctuation-free retrapping
current $I_{\mathrm{r}}^{0}$ is the current value at which the friction must
stop the dynamic state in all cases (this is the property of the model
considered). Similarly, the fluctuations in the low-resistance
`superconducting' state as discussed in the previous subsection for the
unshunted case can allow for premature switching at $0<I<I_{\mathrm{sw}}%
^{0}\equiv I_{\mathrm{c0}}$ but never at $I>I_{\mathrm{sw}}^{0}$ since the
system cannot move through $I_{\mathrm{sw}}^{0}$ without a switch (the bias
current I is assumed to grow linearly in time). One again has an asymmetric
distribution, this time left-skewed or left-tailed. Overall, we can use the
shape of the distribution to gain insights into nature of the state from which
retrapping or switching happens.

In Fig.\ref{Cpow}, a phase diagram for sample S1 is presented which
demonstrates the conditions necessary for the resistive state to be either the
JHNS or a coherent phase dynamics state such as a PSC. The power $P^{\ast
}=I_{\mathrm{W}}V$ at switching and retrapping is calculated by taking the
product of the current through the wire and voltage across the wire at which
the system exhibits switching and retrapping, respectively. Here,
$I_{\mathrm{W}}=I-\frac{V}{R_{\mathrm{S}}}$ , where $I$ is the total current,
which obeys Kirchhoff's Law for current conservation ($I=I_{\mathrm{shunt}%
}+I_{\mathrm{W}}$, where $I_{\mathrm{shunt}}$ is the current through the
shunt). The critical power, $P_{\mathrm{c}}^{\mathrm{\ast}}$, is defined as
the minimum power the wire can sustain and still remain in the JHNS and is
calculated from the power that the unshunted nanowire exhibits at retrapping.
For sample S1, $P_{\mathrm{c}}^{\mathrm{\ast}}$ is calculated to be 0.533 nW
from the unshunted curve in Fig.\ref{pl2}(a). At switching, the unshunted wire
experiences 31 nW of heating, which puts it in the JHNS, where it remains
until the current is reduced below the retrapping current.

With an included shunt, the Joule heating power in the wire at switching is
reduced. For example, when shunted with $75~\mathrm{\Omega}$, the Joule
heating power at switching is 0.359 nW (compared to 31 nW for the unshunted
wire), which is lower than $P_{\mathrm{c}}^{\mathrm{\ast}}$. Thus, the wire
switches to the PSC, which is a superconducting dynamic state (and not the
normal state), and as the current is reduced, it remains in it until
retrapping occurs. Because retrapping occurs from the phase coherent state,
stochastic retrapping is expected for sample S1 when shunted with (from the
experimentally examined values) a $75~\mathrm{\Omega}$ resistor or less. Some
heating is also to be expected since the power at retrapping is still
comparable to $P_{\mathrm{c}}^{\mathrm{\ast}}$.

Guided by the qualitative picture developed above, in the next subsection we
will discuss a model that we will use for simulating the dynamics of the
nanowire. In Subsec. \ref{VIsim} we will address the kinks seen in
Fig.\ref{pl2}(a) and argue that they further support the existence of a
coherent phase dynamic state or a PSC for shunted nanowires.

\subsection{Theoretical model for shunted nanowires}

\begin{figure}[tbh]
\includegraphics[width=9cm]{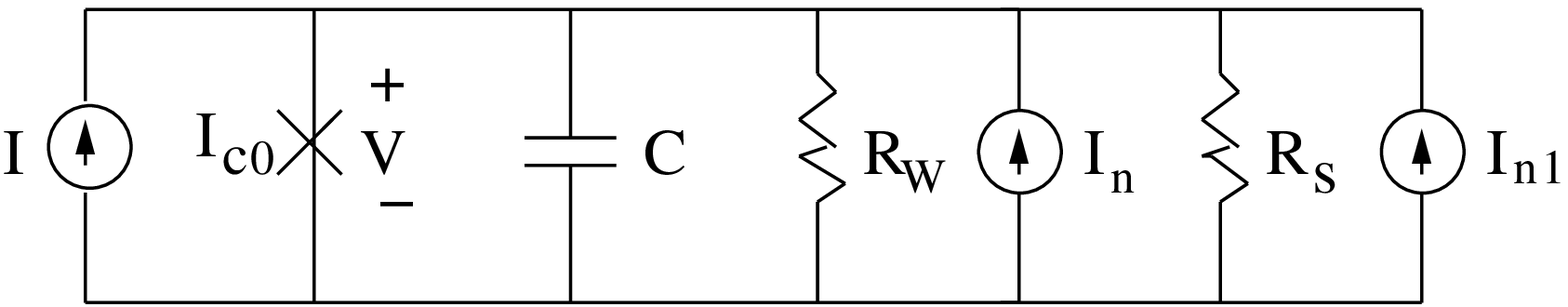}
\caption{Circuit of the resistively capacitively shunted nanowire with an
external shunt resistance $R_{\mathrm{S}}$ and the Johnson-Nyquist
current-noises $I_{\mathrm{n}},~I_{\mathrm{n1}}$.}
\label{RCSJ}
\end{figure}

As argued in the previous subsection, shunting changes the high-resistance
state from JHNS\ to a state with coherent phase dynamics or PSC. When the
value of the shunt resistance is small, we expect that the dynamics of the
nanowire can be modelled by an effectively zero-dimensional circuit model as
is done in the case of a JJ given that the PSC\ behaves like a JJ in series
with the wire. For unshunted nanowires (i.e. infinite shunt value) heating is
important and as discussed above the dynamics of the nanowire is dictated by a
model stemming from a heat diffusion equation. For very large shunt values
some elements of this model might still be important. However, the point of
view we take is to simulate the dynamics of shunted nanowires using a
circuit-element representation and see how well we can reproduce the
experimentally observed behavior.

As discussed in the introduction, the RCSJ model of Stewart and McCumber has
been greatly successful in understanding the physics of JJs. So our goal is to
adopt and extend this model to reflect the experimental set-up and
measurements for the nanowires considered in this paper. The model of Stewart
and McCumber was originally introduced for superconducting JJs to study dc
$V$-$I$ curves displaying hysteresis for light damping \cite{Stewart68,
McCumber68}. This model considered only the time varying phase difference
$\phi(t)$ of the superconducting wave-functions in the weakly coupled
superconductors and neglected any spatial variations of the superconducting
wave-functions and is essentially zero-dimensional. We add an external shunt
resistance in parallel with the superconducting junction in the RCSJ model to
be able to study the behavior for different values of external shunt
resistance as has been studied experimentally. We then simulate the extended
RCSJ model with the Johnson-Nyquist Gaussian white thermal noise coming from
the resistive parts of the circuit. We have not included external noise in our
simulation. Before presenting the details of the model, we pause to provide
some examples of analogies of the observed nanowire behavior with the
established behavior in JJ to further motivate the use of a RSCJ kind of model
for the case of a shunted wire.

The hysteresis in underdamped JJs is due to the bistability of the phase point
in the tilted wash-board potential which depends nonlinearly on the bias
current \cite{Tinkham96}. Damping plays an important role in dictating the
dynamics of the JJ. The experimentally observed saturation of $I_{\mathrm{sw}%
}$ at low shunt values as presented in Section. \ref{Experiment} can be
interpreted to be an effect of high damping (damping $\sim Q^{-1}\sim
R_{\mathrm{S}}^{-1})$ on the premature switching process, and it indicates
that the depairing current is nearly reached for these low values of the
shunt. Another experimental observation we presented earlier is that below
some critical shunt value, the retrapping and switching current become equal
and the hysteresis vanishes. For instance, for sample S1 at 1.8 K at a shunt
value of $10~\mathrm{\Omega}$, the $V$-$I$ curve becomes non-hysteretic as in
Fig. 2(a), and there is no abrupt switch into the resistive state. This can be
interpreted in analogy to JJ\ as follows: At some critical value of the shunt,
the increased damping changes the system from an underdamped junction (with
hysteresis) to an overdamped junction (without hysteresis). In JJs, this
transition occurs when $Q\approx0.84$ \cite{Krasnov07}. the third example of
analogy with JJs is that the mean value of the retrapping current changes
considerably upon changing the value of the shunt resistance. Indeed it is
well-known for JJs that retrapping is very sensitive to the value of damping
and the fluctuation free retrapping current is inversely proportional to the
resistance associated with the JJ.

We give a brief summary of the physics of the RCSJ model in Appendix
\ref{app1} and focus on discussing the details of our extended RCSJ model
below. The displacement current and \textquotedblleft normal\textquotedblright%
\ losses (e.g. quasiparticle tunnel currents) in the nanowire are included in
the model by the shunting capacitance $C$ and resistance $R_{\mathrm{W}}$,
respectively. We also include a Johnson-Nyquist type Gaussian white noise
current source $I_{\mathrm{n}}$ associated with the resistance $R_{\mathrm{W}%
}$ along with the drive current source $I$ \cite{Ambegaokar69, Kurkijarvi70}.
Next we extend the RCSJ model with an external normal resistance
$R_{\mathrm{S}}$ and corresponding Johnson-Nyquist Gaussian white current
noise $I_{\mathrm{n1}}$ for the present experimental set-up of a shunted
nanowire (see Fig.\ref{RCSJ}). Then, the reduced equation of motion for the
phase difference is given by (check Appendix \ref{app1} for a derivation)
\begin{equation}
Q_{\mathrm{0}}^{2}\frac{d^{2}\phi}{dt^{\prime^{2}}}+\big(1+\frac
{R_{\mathrm{W}}}{R_{\mathrm{S}}}\big)\frac{d\phi}{dt^{\prime}}+\sin
\phi=i+i_{\mathrm{n}}+i_{\mathrm{n1}},\label{eom3}
\end{equation}
where $Q_{\mathrm{0}}=(2eI_{\mathrm{c0}}R_{\mathrm{W}}^{2}C/\hbar)^{1/2}$ is
the quality factor, $i=I/I_{\mathrm{c0}}$ is the normalized dc bias current,
$i_{\mathrm{n}}=I_{\mathrm{n}}/I_{\mathrm{c0}},~i_{\mathrm{n1}}=I_{\mathrm{n1}%
}/I_{\mathrm{c0}}$ are the normalized noise currents, $t^{\prime
}=(2eI_{\mathrm{c0}}R_{\mathrm{W}}/\hbar)t$ where $t$ is the physical time
associated with the circuit in Fig.\ref{RCSJ}. Here $I_{\mathrm{c0}}$ is the
fluctuation-free critical current of the nanowire. The time-averaged
steady-state voltage across the wire, $V=I_{\mathrm{c0}}R_{\mathrm{W}}\langle
d\phi/dt^{\prime}\rangle_{t^{\prime}}$, and the noise autocorrelations are
\begin{align}
\langle i_{\mathrm{n}}(t_{1}^{\prime})\rangle &  =0,~~\langle i_{\mathrm{n1}%
}(t_{1}^{\prime})\rangle=0\nonumber\\
\langle i_{\mathrm{n}}(t_{1}^{\prime})i_{\mathrm{n}}(t_{2}^{\prime})\rangle &
=\frac{4ek_{\mathrm{B}}T_{W}}{\hbar I_{\mathrm{c0}}}\delta(t_{1}^{\prime
}-t_{2}^{\prime})~,\label{nprop2}\\
\langle i_{\mathrm{n1}}(t_{1}^{\prime})i_{\mathrm{n1}}(t_{2}^{\prime})\rangle
&  =\frac{4ek_{\mathrm{B}}T_{\mathrm{S}}}{\hbar I_{\mathrm{c0}}}%
\frac{R_{\mathrm{W}}}{R_{\mathrm{S}}}\delta(t_{1}^{\prime}-t_{2}^{\prime
})~,\label{nprop3}
\end{align}
where
$\langle~\rangle$ denotes averaging over the noise realizations (noise
ensemble). The temperature of the nanowire and the shunt resistance are
$T_{\mathrm{W}}$ and $T_{\mathrm{S}}$, respectively. It is possible that the
temperature of the wire is different from that of the shunt resistance, so we
keep here two different noises coming from two different resistances. The
relations in Eqs.(\ref{nprop2},\ref{nprop3}) are known as
fluctuation-dissipation relations. Now if we assume that there is not
significant MQT at the temperature (1.8 K) where the distributions for the
switching current $I_{\mathrm{sw}}$ and the retrapping current $I_{\mathrm{r}%
}$ are measured, then the distributions are due to the thermal fluctuations
arising from the Johnson-Nyquist current noises associated with the resistive
parts of the circuit.

One can write the Eq.(\ref{eom3}) in a little different form as
\begin{equation}
Q^{2}\frac{d^{2}\phi}{dt^{{\prime}^{2}}}+\frac{d\phi}{dt^{\prime}}+\sin
\phi=i+i_{\mathrm{n}}+i_{\mathrm{n1}},\label{eom33}
\end{equation}
where $Q=(2eI_{\mathrm{c0}}R_{\mathrm{T}}^{2}C/\hbar)^{1/2}$, $t^{\prime
}=(2eI_{\mathrm{c0}}R_{\mathrm{T}}/\hbar)t$ and $1/R_{\mathrm{T}%
}=1/R_{\mathrm{W}}+1/R_{\mathrm{S}}$ are respectively the quality factor and
the resistance of the full circuit. But the above form of Eq.(\ref{eom3}) is
more useful in simulations.

\subsection{Shunt dependence of $V$-$I$ characteristics}

\label{VIsim} \begin{figure}[tbh]
\includegraphics[width=8.5cm]{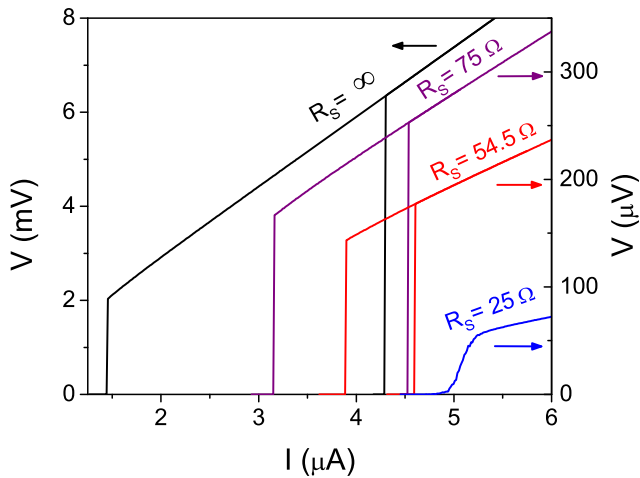}
\caption{Voltage-current characteristics of the nanowire with external shunt
resistance and current noise. The quality factor $Q_{0}=7$ and the
temperatures are related by $T_{\mathrm{W}}=T_{\mathrm{S}}=T \sim1.8~$K and
constant $I_{\mathrm{c0}}=5.55~\mathrm{\mu}$A. The resistance of the unshunted
nano-wire $R_{\mathrm{n}}=1385~ \mathrm{\Omega}$, and the resistance of the
phase-slip center (or shunted nano-wire) $R_{\mathrm{PSC}} \sim0.1
R_{\mathrm{n}}$.}%
\label{IVs}%
\end{figure}Now we simulate Eq.(\ref{eom3}) for $T_{\mathrm{W}}=T_{\mathrm{S}%
}=T_{\mathrm{electrode}}$ to calculate the voltage-current characteristics of
the shunted nanowire in the presence of the current noises from the normal
resistances. In the experiment, one changes the bias current with a finite
current sweep rate, and measures the corresponding voltages. In the
simulation, instead we first fix a bias current and then integrate the above
equations of motion (with suitable initial conditions and current noises) for
a sufficiently long time (this time is the relaxation time or the transient
time), and next calculate the time-averaged voltage by averaging
$d\phi/dt^{\prime}$ over some time interval. For forward current sweep we
choose the initial conditions, $\phi(t^{\prime}=0)=0$,$\frac{d\phi}%
{dt^{\prime}}(t^{\prime}=0)=0$; and for the backward current sweep we use
$\phi(t^{\prime}=0)\neq0$, $\frac{d\phi}{dt^{\prime}}(t^{\prime}=0)\neq0$. The
initial values of $\phi(t^{\prime})$ and $\frac{d\phi}{dt^{\prime}}(t^{\prime
})$ for the backward current sweep can be any finite non-zero values in the
resistive state of the system as after long time of transient dynamics the
exact initial values of $\phi(t^{\prime})$ and $\frac{d\phi}{dt^{\prime}%
}(t^{\prime})$ are irrelevant. We generate the Gaussian white noises
$i_{\mathrm{n}}$ and $i_{\mathrm{n1}}$ at each time step of the simulation
satisfying the noise properties in Eqs.(\ref{nprop2},~\ref{nprop3}) following
the method described in \cite{Allen}.

\begin{figure}[tbh]
\includegraphics[width=8.5cm]{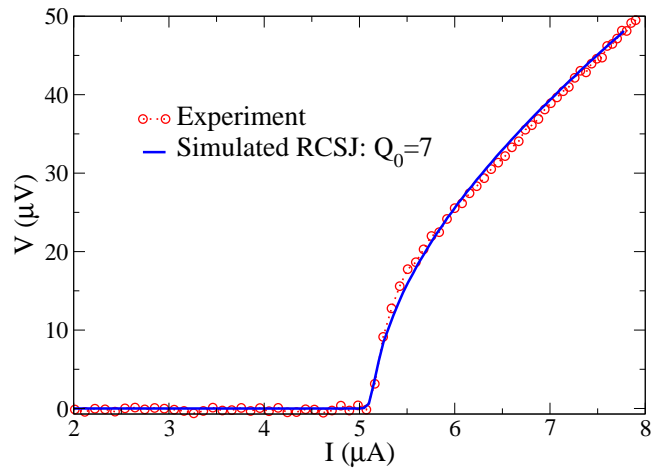}
\caption{Comparison of the voltage-current characteristics of the shunted
nanowire with $R_{\mathrm{S}}=10~\mathrm{\Omega}$ from the experiment and
simulation. We find the best agreement from the simulation of the shunted
nanowire for a quality factor $Q_{\mathrm{0}}=7$ and temperature
$T_{\mathrm{W}}=T_{\mathrm{S}}=T\sim0.13~$K, $I_{\mathrm{c0}}=5.15~\mathrm{\mu
}$A, and a resistance of phase-slip center (or shunted nanowire)
$R_{\mathrm{PSC}}=85~\mathrm{\Omega}$.}%
\label{Rs10}
\end{figure}The voltage-current characteristics are plotted in Fig.\ref{IVs}
for the shunted wires with different values of the normal resistance of the
wire and the shunt resistance. As can be seen from the Eqs.(\ref{eom3}%
,\ref{eom2}), we really do not need explicit values of the resistances in
simulation, but we only need the ratios of the two resistances and a quality
factor. We have checked that for the quality factor $Q_{\mathrm{0}}=7$ and at
a specific temperature $1.8$ K, the transition from hysteretic to
non-hysteretic behavior occurs near the ratio $R_{\mathrm{W}}:R_{\mathrm{S}%
}=4:1$. As discussed in the previous subsection, our present theory with a
coherent phase relationship suits us best to analyze the nanowires shunted
with low shunt resistance, i.e., in the PSC regime. A comparison between
Fig.\ref{pl2}(a) and Fig.\ref{IVs} shows a very good qualitative agreement
between the experimental $V$-$I$ curves and those from simulations. We also
have good quantitative agreement between the experiment and simulation for the
$V$-$I$ curves of the shunted nanowires (PSC regime) with small resistances.
This is shown in Fig.\ref{Rs10} for the shunted nanowire case where
$R_{\mathrm{S}}=10~\mathrm{\Omega}$. We find the best agreement between the
experimental and simulated $V$-$I$ curves for the low shunt resistances when
the effective temperature (related to the thermal noise) of the simulated
nanowires is much lower than the temperature of the superconducting electrodes
in the experiment. Therefore the fluctuation in $I_{\mathrm{sw}}$ due to
thermal noise is expected to be small as we find in experiment. The reason for
such effective thermal noise reduction is not completely clear. It might be
due to the inductance of the shunt resistance. The inductance can cut-off some
of the higher-frequency thermal noise, thus reducing the standard deviation of
the noise current. It also confirms that Joule heating in the shunted
nanowires for lower shunt resistance is greatly reduced compared to the
unshunted case. The resistance of the shunted nanowires $R_{\mathrm{W}}$ which
enters in the simulation through Eqs.\ref{eom3},~\ref{nprop3} is used as a
fitting parameter here. It is necessary to choose it to be much smaller than
the normal resistance of the nanowire $R_{\mathrm{n}}$ to have the best fitted
of $V$-$I$ curves. This is a strong indication that the shunt resistance
drives the nanowire to a phase-coherent PSC state, in which the time-average
supercurrent is not much smaller than the total bias current. Thus we
introduce a new notation for the wire resistance, namely $R_{\mathrm{PSC}}$.
This quantity represents the value of the wire resistance that we have to put
in our model in order to produce the best fits to the experimental $V$-$I$
curves. We find that $R_{\mathrm{PSC}}<<R_{\mathrm{n}}$. For example, we find
$R_{\mathrm{PSC}}=85~\mathrm{\Omega}$ from the simulation for $R_{\mathrm{S}%
}=10~\mathrm{\Omega}$ case. Here we remind that the normal resistance of the
nanowire is $R_{\mathrm{n}}=1385~\mathrm{\Omega}$.

In the experimental $V$-$I$ curve of the shunted nanowire with shunt resistor
25 $\mathrm{\Omega}$ there are kinks which we do not find in simulations. The
kinks can be attributed to the effects of a shunt inductance in series with
the shunt resistor. These kinks are not associated with resonance in the
system because such a resonance would not depend on temperature as these do.
Such inductive effects originate from the fact that the resistor used for
shunting has dimensions of a few centimeters and so has a large inductance
$(\sim20~\mathrm{nH})$. Inductance connected in series with a shunt resistor
is known to cause similar kinks in the $V$-$I$ curves of shunted JJs due to a
complicated dynamic of the phase difference on the junction \cite{Cawthorne98}%
. Thus, the observation of such kinks confirms that the resistive state in our
shunted wires is due to a phase-coherent PSC \cite{Tinkham96} and not due to
Joule heating. Thus we find another indication that by resistively shunting
the nanowire it is possible to change the nature of its resistive state from a
phase-incoherent JHNS to a phase-coherent PSC state.

\subsection{Shunt dependence of switching and retrapping distributions}

\begin{figure}[tbh]
\includegraphics[width=8.5cm]{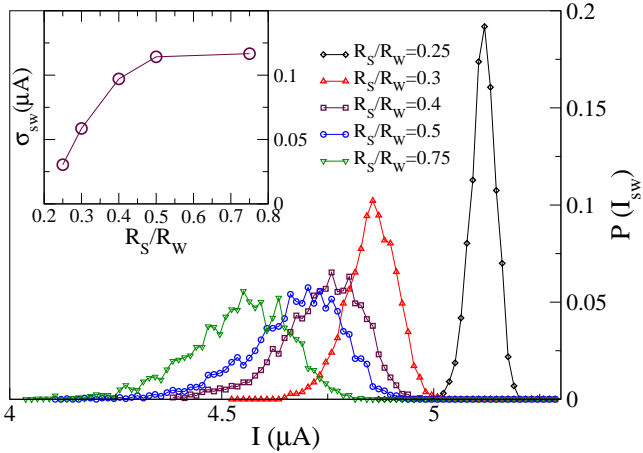}
\caption{Simulated distribution $P(I_{\mathrm{sw}})$ of the switching current
$I_{\mathrm{sw}}$ for different $R_{\mathrm{S}}/R_{\mathrm{W}}$ values at
temperature $T_{\mathrm{W}}=T_{\mathrm{S}}=1.8$ K, constant $I_{\mathrm{c0}%
}=5.55~\mathrm{\mu}$A. The quality factor $Q_{\mathrm{0}}=7$. Inset: Shows the
standard deviation $\sigma_{\mathrm{sw}}$ of switching current distribution as
a function of $R_{\mathrm{S}}/R_{\mathrm{W}}$.}%
\label{sdist1}%
\end{figure}

We first calculate the distributions $P(I_{\mathrm{sw}})$ of the switching
current $(I_{\mathrm{sw}})$ in the shunted nanowire for the same temperature
$T_{\mathrm{W}}=T_{\mathrm{S}}=T_{\mathrm{electrode}}$. We again simulate
Eq.(\ref{eom3}) as before by changing the bias current with zero as initial
values for $\phi$ and $\frac{d\phi}{dt^{\prime}}$, but now we repeat the full
procedure for many realizations of the thermal noise. We count a switch from
the metastable to the running state when the nanowire spends more than half
the time in the running state over some sufficiently long time period
($\tau_{\mathrm{s}}$) of simulation. This gives us a distribution for
$I_{\mathrm{sw}}$ which is plotted in Fig.\ref{sdist1} for the shunted wire.
We find a good qualitative agreement between the simulation and experiment for
the switching distributions of the shunted nanowires as shown in
Fig.\ref{pEdist} and Fig.\ref{sdist1}. In the inset of Fig.\ref{sdist1} we
show the standard deviation $\sigma_{\mathrm{sw}}$ of the simulated
$P(I_{\mathrm{sw}})$ with $R_{\mathrm{S}}/R_{\mathrm{W}}$ and it matches with
the trend of the standard deviation of the measured $P(I_{\mathrm{sw}})$ from
the experiment. Here we mention that $\sigma_{\mathrm{sw}}$ shows a
non-monotonic behavior with $R_{\mathrm{S}}/R_{\mathrm{W}}$ in the RCSJ model
for higher values of $R_{\mathrm{S}}$, for example, $R_{\mathrm{S}%
}>R_{\mathrm{W}}$ at a constant temperature. This non-monotonic behavior of
$\sigma_{\mathrm{sw}}$ in the RCSJ model with $R_{\mathrm{S}}/R_{\mathrm{W}}$
at a constant temperature is similar to non-monotonic behavior of
$\sigma_{\mathrm{sw}}$ with temperature for a constant $R_{\mathrm{S}%
}/R_{\mathrm{W}}$. It will be discussed in the last subsection. However,
experimental results show the value of $\sigma_{\mathrm{sw}}$ is greater for
the unshunted case $(R_{\mathrm{S}}=\infty)$ than the shunted case. This also
indicates that the unshunted nanowire is not in a coherent PSC state but is
dominated by Joule heating which increases the effective temperature of the wire.

\begin{figure}[tbh]
\includegraphics[width=8.5cm]{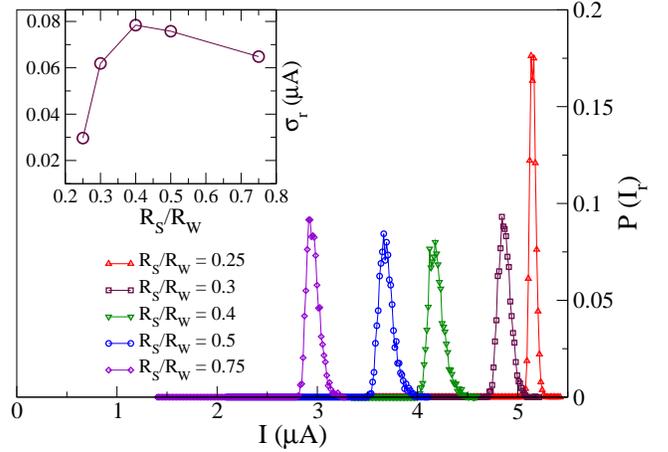}
\caption{Simulated distribution $P(I_{\mathrm{r}})$ of the retrapping current
$I_{\mathrm{r}}$ for different $R_{\mathrm{S}}/R_{\mathrm{W}}$ values at
temperatures $T_{\mathrm{W}}=T_{\mathrm{S}}=1.8~$K, constant $I_{\mathrm{c0}%
}=5.55~\mathrm{\mu A}$. The quality factor $Q_{\mathrm{0}}=7$. Inset: Shows
the standard deviation $\sigma_{\mathrm{r}}$ of the retrapping current
distribution as a function of $R_{\mathrm{S}}/R_{\mathrm{W}}$.}%
\label{sdist2}%
\end{figure}

We next simulate the extended RCSJ model to understand the measured
distributions $P(I_{\mathrm{r}})$ of the retrapping current for different
shunt resistances. Here we choose a fixed temperature. The simulation method
is similar to finding the switching distributions, but now we start from a PSC
state with nonzero initial conditions for $\phi(0)$ and $\frac{d\phi
}{dt^{\prime}}(0)$. We reduce the bias current and count a retrapping event
from the running to metastable state when the nanowire spends less than half
of the time in the running state over time period of $\tau_{\mathrm{s}}$. The
simulated retrapping distributions are plotted in Fig.\ref{sdist2} for
different $R_{\mathrm{S}}/R_{\mathrm{W}}$ values at a constant temperature. We
find from Fig.\ref{rpEdist} that the standard deviation of the retrapping
current distributions falls slightly (within the experimental noise limit)
with decreasing shunt resistance. The standard deviation of the simulated
retrapping current distribution also falls slightly below $R_{\mathrm{S}%
}/R_{\mathrm{W}}\sim0.4$ for $Q_{\mathrm{0}}=7$ which is similar to the
experiment. But we also find from the simulation of the RCSJ model that the
standard deviation of $P(I_{\mathrm{r}})$ decays slightly with increasing
shunt resistance above $R_{\mathrm{S}}/R_{\mathrm{W}}\sim0.4$ for
$Q_{\mathrm{0}}=7$, but it never goes to zero at higher shunt resistances for
the RCSJ model with coherent dynamics. The standard deviation of the measured
retrapping current distribution for the unshunted wire is almost zero within
the noise limit; this is consistent with the existence of JHNS in the
unshunted nanowire and PSCs in shunted wires. The width of the simulated
switching and retrapping distributions are much greater than the experimental
results at the same temperature $T=1.8$ K. This might be again due to the
inductance of the shunt resistor which can effectively reduce the thermal
noise in the system.

\subsection{Temperature dependence: shunted vs. unshunted nanowires}

\begin{figure}[tbh]
\includegraphics[width=8cm]{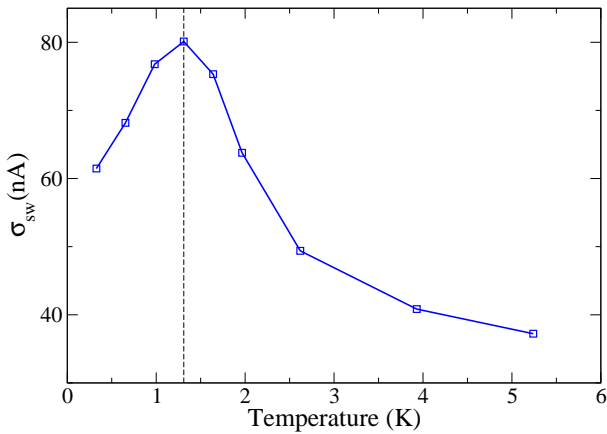}
\caption{ The standard deviation $\sigma_{\mathrm{sw}}$ vs. $T$ for the
simulated switching events in the RCSJ model with $Q_{\mathrm{0}}=7$ and
$R_{\mathrm{S}}/R_{\mathrm{W}}=0.3$. }%
\label{swwd}
\end{figure}Finally we simulate the extended RCSJ model with an external low
shunt resistance to find the temperature dependence of the standard deviation
$\sigma_{\mathrm{sw}}$ of switching current. We use the same scheme as
previous sub-sections to determine the switching current in the simulation at
different temperatures. We plot $\sigma_{\mathrm{sw}}$ vs. $T$ in
Fig.\ref{swwd} for $R_{\mathrm{S}}/R_{\mathrm{W}}=0.3$ and $Q_{\mathrm{0}}=7$.
We find that temperature dependence of simulated $\sigma_{\mathrm{sw}}$ due to
the thermal fluctuations is non-monotonic just as in the experiment (see
Fig.\ref{swwd1}) for the shunted nanowires. A non-monotonic temperature
dependence of $\sigma_{\mathrm{sw}}$ due to the thermal fluctuations has also
been obtained previously for various JJs \cite{Krasnov05, Mannik05, Krasnov07}
and our theoretical analysis not only highlights its ubiquity but also
provides a way of obtaining it in terms of a RCSJ\ kind of modelling. We also
find a non-monotonic temperature dependence of the standard deviation of
retrapping current in our simulation. In our numerical study here we only
consider effect of thermal fluctuations in phase-slips, thus one needs to go
beyond the present study to include effect of macroscopic quantum tunneling on
the temperature dependence of $\sigma_{\mathrm{sw}}$ in the fully quantum
regime. 

In unshunted nanowires too it has been shown that the standard deviation of
the switching current distribution is non-monotonic as a function of
temperature \cite{Shah08, Pekker09}. In the thermal regime at higher
temperatures $T>T^{\mathrm{\ast}}$ multiple phase slips are required before
the wire switches to the normal state and the standard deviation increases as
the temperature is decreased \cite{Sahu09}. At slightly lower temperatures
$T<T^{\mathrm{\ast}}$ a single thermal phase slip causes the wire to switch to
the normal state and the standard deviation decreases with a decrease in
temperature. At low temperatures when QPS are present, depending on how
$T^{\mathrm{\ast}}$ compares with the temperature of crossover from TAPS to
the QPS, one can get different behaviors. With an applied external shunt, the
increased dissipation is expected to decrease the temperature at which the
crossover from thermal activation to MQT takes place.

\section{Concluding remarks}

\label{Conclusion}

We have undertaken a detailed study of the effect of external resistance
shunts on the behavior of superconducting nanowires. Shunting has a strong
effect on the behavior of the nanowire. We find that the statistics of the
switching and retrapping currents significantly depends on the value of the
shunt resistance. The temperature dependence of the mean switching current in
strongly shunted nanowires is consistent with the Bardeen prediction for the
temperature dependence of the critical current \cite{Bardeen62}; this
indicates that the switching current can be controllably driven very near to
the depairing current through external resistive shunting. The retrapping
current on the other hand increases and becomes more stochastic, at least for
moderate shunting. We demonstrate that the shunting, even with a large
resistance value, can be used to control the phase slip events in the wire. We
suggest a model based on the Stewart-McCumber RCSJ\ model, which is
generalized to include two resistive elements, corresponding to the effective
resistance of the wire (with a phase slip center), and the resistance of the
shunt. The model provides a semi-quantitative description to the data.
Moreover, it provides insights into developing a circuit-element
representation of a superconducting nanowire. 

Our work opens up many interesting avenues towards developing a fundamental
understanding and control of coherence and dissipation in nanowires as well as
its relation to possible quantum phase transitions. It will be important to
develop a model that incorporates heating as well as coherent dynamics and
studies the entire crossover in going from unshunted nanowires (i.e. infinite
shunt resistance in parallel) where heating is most important to the case of
very low shunt resistance values where heating is least important. It would be
extremely interesting and relevant also to explore the low temperature quantum
regime in more detail and to study the implications of our work for quantum
computing and other technological applications of nanowires. As an example,
nanowires have been used as photon counters \cite{Kerman06}, which are
important in radioastronomy. The switching events studied in this paper,
represent so-called \textquotedblleft dark counts\textquotedblright, in the
terminology of the photon detection community. Understanding the physics of
dark counts is important for the purpose of improving superconducting photon
detectors. The fact that the standard deviation of switching current becomes
smaller with the inclusion of a shunt resistor has relevance to photon
detectors, since dark counts can be reduced by shunting. 

\section{Acknowledgments}

The experimental material is based upon work supported by the U.S. Department
of Energy, Division of Materials Sciences under Award No. DE-FG02-07ER46453,
through the Frederick Seitz Materials Research Laboratory at the University of
Illinois at Urbana-Champaign. Part of the experimental work was carried out in
the Frederick Seitz Materials Research Laboratory Central Facilities,
University of Illinois. DR and NS acknowledge the University of Cincinnati for
financial support while developing the theoretical understanding of the
experiment. NS\ acknowledges the hospitality of Tata Institute of Fundamental
Research while working on the manuscript of the paper. Finally, the authors
would like to thank Myung-Ho Bae and M. Sahu for fruitful discussions.

\appendix

\section{Resistively and capacitively shunted Josephson junction (RCSJ) model}

\label{app1}

\begin{figure}[tbh]
\includegraphics[width=3.0in]{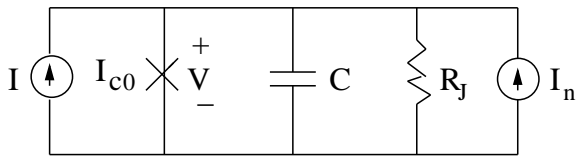}
\caption{Stewart-McCumber resistively and capacitively shunted Josephson
junction (RCSJ) model with the Johnson-Nyquist thermal current noise.}%
\label{RCSJ1}%
\end{figure}Here we briefly digress the main features of the RCSJ model (see
Fig.\ref{RCSJ1}) introduced by Stewart and McCumber \cite{Stewart68,
McCumber68}. We find the equation of motion for the time varying phase
difference $\phi(t)$ of the superconducting wave-functions of the circuit in
Fig.\ref{RCSJ1} by applying the well-known Josephson dc and ac
relations~\cite{Josephson1,Josephson2,Josephson3} for the current-phase
$(I-\phi)$ and the voltage-phase $(V-\phi)$,
\begin{align}
I_{\mathrm{J}} &  =I_{\mathrm{c0}}\sin\phi~,\label{JR1}\\
\frac{d\phi}{dt} &  =\frac{2eV}{\hbar}~,\label{JR2}%
\end{align}
where $I_{\mathrm{c0}}$ is the fluctuation-free intrinsic critical current of
the junction. The equation of motion for $\phi(t)$ of the circuit in
Fig.\ref{RCSJ1} is given by,
\begin{equation}
\frac{C\hbar}{2e}\frac{d^{2}\phi}{dt^{2}}+\frac{\hbar}{2eR_{\mathrm{J}}}%
\frac{d\phi}{dt}+I_{\mathrm{c0}}\sin\phi=I+I_{\mathrm{n}},\label{eom1}
\end{equation}
along with the Gaussian white noise properties for the Johnson-Nyquist thermal
current noise $I_{\mathrm{n}}$,
\begin{align}
\langle I_{\mathrm{n}}(t)\rangle &  =0\nonumber\\
\langle I_{\mathrm{n}}(t_{1})I_{\mathrm{n}}(t_{2})\rangle &  =\frac
{2k_{\mathrm{B}}T_{\mathrm{J}}}{R_{\mathrm{J}}}\delta(t_{1}-t_{2}%
)~,\label{nprop1}%
\end{align}
where $T_{\mathrm{J}}$ is the temperature of the JJ with capacitance $C$ and
resistance $R_{\mathrm{J}}$. The resistance $R_{\mathrm{J}}$ measures
dissipation in the JJ in the finite voltage regime, without affecting the
lossless dc zero voltage regime, and $C$ indicates the geometric shunting
capacitance between the two superconducting electrode \cite{Tinkham96}.
The Eq.(\ref{eom1}) of motion for the junction phase can be rewritten in terms
of dimensionless parameters as
\begin{equation}
Q_{\mathrm{0}}^{2}\frac{d^{2}\phi}{dt^{{\prime}^{2}}}+\frac{d\phi}{dt^{\prime
}}+\sin\phi=i+i_{\mathrm{n}},\label{eom2}
\end{equation}
where $Q_{\mathrm{0}}=(2eI_{\mathrm{c0}}R_{\mathrm{J}}^{2}C/\hbar)^{1/2}$ is
the quality factor of the linearized equation of motion, $i=I/I_{\mathrm{c0}}%
$, $i_{\mathrm{n}}=I_{\mathrm{n}}/I_{\mathrm{c0}}$, and $t^{\prime
}=(2eI_{\mathrm{c0}}R_{\mathrm{J}}/\hbar)t$ is the normalized time.
The term $\frac{d\phi}{dt^{\prime}}$ is damping as it breaks the
time-reversibility of the equation and introduces dissipation. The strength of
damping is proportional to $1/R_{\mathrm{J}}$ and is inversely related to the
quality factor. In this notation, the time-averaged steady-state voltage
across the junction, $V=I_{\mathrm{c0}}R_{\mathrm{J}}\langle d\phi/dt^{\prime
}\rangle_{t^{\prime}}$. The noise correlation in the scaled time, $\langle
i_{\mathrm{n}}(t_{1}^{\prime})i_{\mathrm{n}}(t_{2}^{\prime})\rangle
=(4ek_{\mathrm{B}}T_{\mathrm{W}}/\hbar I_{\mathrm{c0}})\delta(t_{1}^{\prime
}-t_{2}^{\prime})$. The usual McCumber parameter $\beta_{\mathrm{c}%
}=Q_{\mathrm{0}}^{2}$ and Stewart parameter $\omega_{\mathrm{0}}%
\tau=Q_{\mathrm{0}}$ with $\omega_{\mathrm{0}}=\sqrt{2eI_{\mathrm{c0}}/\hbar
C}$ and $\tau=R_{\mathrm{J}}C$. For the circuit in Fig.\ref{RCSJ} we replace
$R_{\mathrm{J}}$ in Eq.(\ref{eom1}) by $R_{\mathrm{T}}$ where $1/R_{\mathrm{T}%
}=1/R_{\mathrm{W}}+1/R_{\mathrm{S}}$. Then we derive either Eq.(\ref{eom3}) or
Eq.(\ref{eom33}) following the similar steps to get Eq.(\ref{eom2}) from
Eq.(\ref{eom1}).

In the absence of thermal current noise $I_{\mathrm{n}}$ at zero temperature
(also neglecting quantum fluctuations), the zero-voltage state or 0 state is
stable at all bias levels less than the ideal critical current ($|i|<1$), and
the voltage state or 1 state is stable at all bias levels greater than a
minimum value designated by a fluctuation free re-trapping current
$i_{\mathrm{r0}}$. The value of $i_{\mathrm{r0}}~(\equiv I_{\mathrm{r0}%
}/I_{\mathrm{c0}})$ is determined entirely by the quality factor $Q_{0}$ and
decreases with increasing $Q_{0}$ as a smaller tilt is sufficient to support
the running (finite voltage) state when damping is less. For $Q_{0}<0.8382$,
the damping is sufficient that a running state is not possible unless the
potential decreases monotonically and in this case $i_{\mathrm{r0}}=1$. For
$Q_{0}>0.8382$, a running state is possible even when the potential has local
minima. \cite{Kautz90} In this case $i_{\mathrm{r0}}<1$ and the $V$-$I$ curve
is hysteretic. In the limit of large $Q_{0}$, $i_{\mathrm{r0}}=4/\pi Q_{0}$
($Q_{0}>3$). \cite{Stewart68, Kautz90}

The phase dynamics described in Eq.(\ref{eom2}) can be visualized as the
damped motion of a Brownian particle in the tilted washboard potential
$U(\phi)=-(i\phi+\cos\phi)$. In the under-damped regime $Q_{0}>1$, the
zero-voltage state and the resistive state correspond to the particle trapped
by the energy barrier $\Delta U$ and running downward along the tilted
potential, respectively. Escape from the potential (0 state to 1 state) can
occur even for $i<1$ due to the thermal and the quantum fluctuations.

\end{document}